\documentclass[%
 reprint,%
 amssymb, amsmath,%
 aip,cha,%
]{revtex4-1}

\usepackage{docs}%
\usepackage{bm}%
\usepackage[colorlinks=true,linkcolor=blue]{hyperref}%
\usepackage{graphicx}
\expandafter\ifx\csname package@font\endcsname\relax\else
 \expandafter\expandafter
 \expandafter\usepackage
 \expandafter\expandafter
 \expandafter{\csname package@font\endcsname}%
\fi
\begin{document}

\title{Two-dimensional magnetohydrodynamic turbulence with large and small
energy-injection length scales}

\author{Debarghya Banerjee}%
\email{debarghya.banerjee@ds.mpg.de}
\affiliation{ Max Planck Institute for Dynamics and Self-Organization, Am Fa\ss berg 17, 37077 G\"ottingen, Germany. }%

\author{Rahul Pandit}
\email{rahul@iisc.ac.in; also at Jawaharlal Nehru Centre for Advanced Scientific Research, Bangalore, India}
\affiliation{Centre for Condensed Matter Theory, Department of Physics,
		Indian Institute of Science, Bangalore 560012, India.}

\date{\today}%

\begin{abstract}

Two-dimensional magnetohydrodynamics (2D MHD), forced at (a) large length
scales or (b) small length scales, displays turbulent, but
statistically steady, states with widely different statistical
properties. We present a systematic, comparative study of these two
cases (a) and (b) by using direct numerical simulations (DNSs). We find
that, in case (a), there is energy equipartition between the magnetic
and velocity fields, whereas, in case (b), such equipartition does not
exist.  By computing various probability distribution functions (PDFs),
we show that case (a) displays extreme events that are much less common
in case (b).

\end{abstract}

\maketitle


\section{Introduction}
\label{intro}

Statistically steady turbulence, in fluids or magnetohydrodynamics (MHD), is
sustained by external forcing. Away from boundaries and the scales at which
energy is injected into the fluid, such turbulence is statistically homogeneous
and isotropic. The length scale at which the external forcing acts determines,
to a large extent, the nature and statistical properties of this
turbulence~\cite{Biskamp2003,Verma2004}. Nonlinear cascades of the energy or
other quadratic invariants (e.g., the enstrophy $\Omega \equiv \int d^3 {\bf
r} ~ |\nabla \times {\bf u}|^2$, or the magnetic helicity $H_M = \int d^3 {\bf r}
~{\bf a} \cdot {\bf b}$, where ${\bf u}$ and ${\bf b}$ are, respectively, the
velocity and magnetic fields, and ${\bf a}$ is the magnetic vector potential)
lead to different inertial ranges of length scales that lie between $L$, the
typical linear system size, and $\eta$, the length scale at which dissipation
becomes significant. In a turbulent fluid, the Fourier-space energy spectrum
$E(k) \sim k^{\gamma_1}$, where $k$ is the wave number and $\gamma_1$ an
exponent that characterizes this scaling form in the inertial range $2 \pi / L
\ll k \ll 2 \pi / \eta$.  In MHD turbulence, similar scaling forms hold for
both the magnetic- and fluid-energy spectra~\cite{Biskamp2003,Verma2004}
$E^b(k)$ and $E^u(k)$, respectively.  Two-dimensional (2D) fluid turbulence
displays two inertial
ranges~\cite{Frisch1995,Kraichnan1967,Fjortoft1953,Batchelor1969,Leith1972}: If
the energy-injection or forcing length scale is $2 \pi / k_{\rm inj}$, there
are two scaling regimes in $E(k)$, namely, the \textit{forward-cascade regime},
for $k_{\rm inj} \ll k \ll 2 \pi/ \eta$, in which the enstrophy cascades from
the forcing length scales towards the dissipation scale, and (b) the
\textit{inverse-cascade regime}, for $2 \pi / L \ll k \ll k_{\rm inj}$, in
which the energy goes from the injection length scale towards larger length
scales~\cite{Frisch1995,Kraichnan1967,Fjortoft1953,Batchelor1969,Leith1972,PanditOverview17}.
In 2D MHD turbulence, 2D-fluid-turbulence-type arguments hold, but there is a
forward cascade of energy and an inverse cascade of magnetic
helicity~\cite{Frisch1975,PanditOverview17,Banerjee2014}, if we assume that the
cross helicity $H_C = \int d^3 {\bf r}~ {\bf u} \cdot {\bf b}$ has a negligible
effect on the dynamics. Other examples of inverse cascades can be found in
turbulence in quasi-geostrophic flows~\cite{Bistagnino2008,Bernard2007}, in
rotating fluids~\cite{Smith1999}, in fluid films with added
polymers~\cite{Gupta2015}, and in 3D MHD, which has an inverse cascade of
magnetic helicity and forward cascades of the energy and cross
helicity~\cite{Alexakis2006,Biskamp2003}.  By contrast, 3D fluid turbulence shows
no inverse cascade, but only a forward cascade of energy.

The direction of cascades in turbulence can be predicted by using arguments of
equilibrium statistical physics~\cite{Kraichnan1980}. If we consider the invariants of the
system to have a Gibbsian distribution, and we calculate the spectra of
these invariants, then a maximum in the spectrum at small (large) $k$ indicates
an inverse (forward) cascade.
This was first shown, for 2D fluid turbulence, in the seminal work of 
Kraichnan ~\cite{Kraichnan1980}, who proposed the inverse-energy cascade, which 
implied the formation of large-scale vortical structures. These
predictions~\cite{Kraichnan1980} were based on arguments of equilibrium
statistical physics~\cite{Lee1952} applied to the 2D, Galerkin-truncated Euler
equations, whose finite-dimensional phase space allowed the
Galerkin-truncated system to thermalise to an equilibrium state. This
statistical-mechanical technique has been used, subsequently, to predict the
natures of cascades in a wide variety of turbulent systems~\cite{Krstulovic2009}, including
MHD turbulence: it has revealed the inverse cascades of (a) the magnetic helicity in
3D MHD turbulence and (b) the squared magnetic potential $ A \equiv \Sigma_k |\psi(k)|^2$
in 2D MHD turbulence~\cite{Banerjee2014}. From the forward cascade of total energy
$E^u(k) + E^b(k) = \Sigma_k 1/2(|u(k)|^2 + |b(k)|^2)$) and the inverse cascade of 
$A$, and by using dimensional analysis, it is possible to 
predict the scaling forms of the energy spectrum in the forward- and
inverse-cascade regimes~\cite{Biskamp2003} in 2D MHD turbulence.

In Ref.~\cite{Banerjee2014}, we have outlined the dimensional arguments that
are used for extracting the scaling exponents in the inverse-cascade regime of
2D MHD turbulence.  We give below similar arguments for the forward-cascade
regime:
\begin{eqnarray}
\lbrack u \rbrack &=& \frac{L}{T} ;\, 
\lbrack b \rbrack = \frac{L}{T} ; \nonumber \\
\lbrack k \rbrack &=& \frac{1}{L}  ;\nonumber \\
\lbrack \epsilon_{u} \rbrack &=& \frac{L^2}{T^3} ;\,
\lbrack \epsilon_{b} \rbrack = \frac{L^2}{T^3} ; \nonumber \\
\lbrack |u(k)|^2 \rbrack &=& \frac{L^3}{T^2} ;\,
\lbrack |b(k)|^2 \rbrack = \frac{L^3}{T^2}; 
\label{eq:dimensions1}
\end{eqnarray}
here, we indicate by square brackets the dimensions of different
quantities and express them as powers of length $L$ and time $T$.
(Recall that the velocity and magnetic fields have the same units in the standard
formulation of MHD; and $\epsilon_u$ and $\epsilon_b$ are the dissipation rates
of kinetic energy and magnetic energy, respectively.) We use the type of power-law
Ansatz employed by Kolmogorov~\cite{K41} in 1941 (K41) for 3D fluid turbulence,
namely,
\begin{equation}
E(k) \sim \epsilon^{\gamma_1} k^{\gamma_2};
\label{eq:dimensions2}
\end{equation}
by dimensional analysis we obtain
\begin{equation}
\frac{L^3}{T^2} = \left( \frac{L^2}{T^3} \right)^{\gamma_1} \left( \frac{1}{L} \right)^{\gamma_2},
\end{equation}
and thence $\gamma_1=2/3$ and $\gamma_2=-5/3$, i.e.,
\begin{equation}
E(k) \sim \epsilon^{2/3} k^{-5/3}.
\end{equation}
Note that these dimensional and scaling arguments are predicated upon a K41-type
phenomenology; strictly speaking this is not correct because of intermittency
corrections that lead to multifractality~\cite{Frisch1995}; furthermore, these
arguments do not account for a bottleneck in the energy spectrum at intermediate
wavenumbers~\cite{Frisch2008,Frisch2013}. For discussions of energy-spectral
exponents in 3D MHD turbulence (the K41 $-5/3$ versus the Iroshnikov-Kraichnan
$-3/2$), we refer the reader to
Refs.~\cite{Biskamp2003,Verma2004,Gibbon2016,Biskamp2000,Muller2000,Mininni2007,Mininni2009,Sahoo2011,Basu2018}.

Some recent studies ~\cite{Seshasayanan2014,Seshasayanan2016} have examined the
transition from an inverse to a forward cascade in 2D MHD turbulence as a
function of the forcing. We carry out a systematic comparison of the properties
of statistically steady, homogeneous and isotropic 2D MHD turbulence forced at
(a) large length scales and (b) small length scales, by using direct numerical
simulations (DNSs).  We show that there is energy equipartition between the
magnetic and velocity fields in case (a) but not in case (b).  By computing
various probability distribution functions (PDFs), we show that case (a)
displays extreme events that are much less common in case (b).


The PDFs of the vorticity $\omega$, the current density $j$, and the 2D analog
of the magnetic vector potential $\psi$ and the stream function $\phi$ (see
below) deviate from a Gaussian PDF in turbulent flows. However, it has been
noted~\cite{Boffetta2000} that, in inverse-cascade regimes, the deviations from
Gaussian PDFs are much less than in the forward-cascade regime. Furthermore,
the PDF of the Okubo-Weiss parameter~\cite{okubo,weiss} helps us to quantify
the dominance of vortical regions over strain-rate-dominated regions in 2D
flows~\cite{perlekarnjp}. This parameter has also been used to examine polymer
stretching in a turbulent 2D fluid with polymer additives~\cite{Gupta2015}. The
2D MHD analogs of the Okubo-Weiss parameter have been introduced in
Refs.~\cite{Banerjee2014,shivamoggi}. The PDF of the cosine of the angle
between the velocity and the magnetic field in 2D MHD
turbulence~\cite{Banerjee2014} and can be used to estimate the importance of
alignment-induced suppression of the nonlinear terms in the induction equations
(see below). We quantify the differences between PDFs of such quantities for
cases (a) and (b).

The remaining part of this paper is organised as follows. In the next section
(Sec.~\ref{sec:eq}) we present the equations and numerical methods we use. This
is followed by a section on our results (Sec.~\ref{sec:res}). We end in
Sec.~\ref{sec:concl} with conclusions.

\section{Equations and numerical methods}
\label{sec:eq}

We write the 2D MHD equations in the following vorticity-stream-function form\cite{Banerjee2014}:
\begin{eqnarray}
\nonumber
\frac{\partial \omega}{\partial t} + {\bf u} \cdot {\bf \nabla} \omega + \mu^{\omega} \omega &=& -\nu \nabla^4 \omega + f^{\omega} + {\bf b} \cdot {\bf \nabla} j, \\
\nonumber
\frac{\partial \psi}{\partial t} + {\bf u} \cdot {\bf \nabla} \psi + \mu^{\psi} \psi &=& -\eta \nabla^4 \psi + f^{\psi}; \\
\label{eq:2dmhd}
\end{eqnarray}
here, the magnetic field $\bf b$ and the velocity field $\bf u$ are related to
the (2D) magnetic vector potential $\psi$ and the stream function $\phi$ via
${\bf b} = \hat{z} \times {\bf \nabla} \psi$ and ${\bf u} = \hat{z} \times {\bf
\nabla} \phi$, with $\hat{z}$ the unit normal to our 2D domain; furthermore, $j
= \nabla^2 \psi$ and $\omega = \nabla^2 \phi$. This form of the 2D MHD
equations ensures that the incompressibility condition $\nabla \cdot {\bf u}$
$=0$ and $\nabla \cdot \bf{ b}$ $= 0$ are satisfied. We use second-order
hyperviscosity $\nu$ and magnetic hyperdiffusivity $\eta$, with a squared
Laplacian, instead of the conventional viscosity and diffusivity to attain
extended scaling ranges in the energy spectra of the  statistically steady
turbulent state of 2D MHD turbulence. [High-order hyperviscosity enhances the
bottleneck in the energy spectrum~\cite{Frisch2008,Frisch2013}, leads to an
effective Galerkin truncation that can, in turn, result in thermalization;
therefore, we restrict ourselves to second-order hyperviscosity and magnetic
hyperdiffusivity.]  The coefficients of friction are  $\mu^{\omega}$ and
$\mu^{\psi}$; and the forcing terms are:
\begin{eqnarray} 
f^{\omega} &=& -f_{\rm amp}^{\omega} k_{\rm inj}\cos(k_{\rm inj} x); \nonumber \\ 
f^{\psi} &=& f_{\rm amp}^{\psi} \frac{1}{k_{\rm inj}}\cos(k_{\rm inj} y).  
\end{eqnarray} 
Thus, $k_{\rm inj}$ is the wave number at which we inject energy into the
system.

We employ the pseudospectral method~\cite{canuto} for our DNSs, in a 2D,
square, simulation domain (side $L = 2\pi$ and periodic boundary conditions), and
the $2/3$ dealiasing method. We use a second-order, Runge-Kutta method for
time marching. In addition to the spatiotemporal evolution of $\omega$ and
$\psi$, we obtain $\bf{u}, \, \bf{b}, \, \phi$, and $j$.  The fluid Reynolds
number is  $Re = v_{\rm rms} 2 \pi/\nu_{\rm eff}$, its magnetic analog is $Re_M
= v_{\rm rms} 2 \pi/\eta_{\rm eff}$, the root-mean-square velocity is $v_{\rm
rms} = \sqrt{E^u}$, and the effective viscosity and magnetic diffusivity (subscript eff) 
are, respectively,
\begin{eqnarray}
\nu_{\rm eff} &=& \frac{\sum_k \nu k^{2\alpha} E^u(k)}{\sum_k  k^{2} E^u(k)} , \nonumber \\
\eta_{\rm eff} &=& \frac{\sum_k \eta k^{2\alpha} E^b(k)}{\sum_k  k^{2} E^b(k)},
\end{eqnarray}
the box-size eddy turnover time is $\tau_{\rm eddy} = 2\pi/u_{\rm rms}$,
and the kinetic- and magnetic-energy spectra are
$E^u(k) = \Sigma_{{\bf k} \ni |{\bf k}| =k}
|{\bf u (k)}|^2$ and $E^b(k) = \Sigma_{{\bf k} \ni |{\bf k}| =k} |{\bf b
(k)}|^2$, respectively. 

\begin{table*}
\begin{center}
\begin{tabular}{l c c c c c c c c c}
\hline
Runs & $N$ & $\nu=\eta$ & $\mu^{\omega}$ & $\mu^{\psi}$ & $k_{\rm inj}$ & $f_{\rm amp}^{\omega}$ & $f_{\rm amp}^{\psi}$ & $\tau_{\rm eddy}$ & $\tau_{\rm av}$ \\ \hline
R1 & $1024$ & $10^{-8}$ & $0.1$ & $0.1$ & $2$ & $0.1$ & $0.01$ & $20$ & $100$ \\ 
R2 & $1024$ & $10^{-8}$ & $0.0$ & $0.05$ & $250$ & $0.01$ & $0.001$ & $10$ & $100$\\ \hline
\end{tabular}
\end{center}
\caption{Parameters for our two DNS runs R1 and R2, which use $N^2$ 
collocation points, kinematic hyperviscosity $\nu$ and magnetic hyperdiffusivity 
$\eta$ (see text), the friction coefficients $\mu^{\omega}$ and $\mu^{\psi}$, 
energy-injection wave number $k_{\rm inj}$, and forcing amplitudes 
$f_{\rm amp}^{\omega}$ and $f_{\rm amp}^{\psi}$.} 
\label{table:para}
\end{table*}

\section{Results}
\label{sec:res}

We compare some statistical properties of 2D MHD turbulence, for which we
obtain statistically steady states, from our DNSs with forcing such that there
are two different energy-injection scales. In particular, our two DNSs are
distinguished by $k_{\rm inj}$, the wavenumber at which we inject energy into
the system. In our first DNS (run R1), $k_{\rm inj}=2$ and, in the second (run
R2), $k_{\rm inj}=250$. We show that various statistical properties of the
turbulent states, in the runs R1 and R2, are strikingly different. We establish
this by calculating  and comparing, for these two runs, (a) the time evolution
of the kinetic, magnetic, and total energies, (b) energy spectra, (c)
probability distribution functions (PDFs) of the vorticity, current density,
fluid stream function, magnetic potential, of the cosine of the angle between
the velocity and magnetic fields, and of the Okubo-Weiss
parameter~\cite{okubo,weiss} and its magnetic
analog~\cite{Banerjee2014,shivamoggi}, which help us to characterise the
topology of the flow. 

\begin{figure}[htbp!]
\includegraphics[width=0.95\columnwidth]{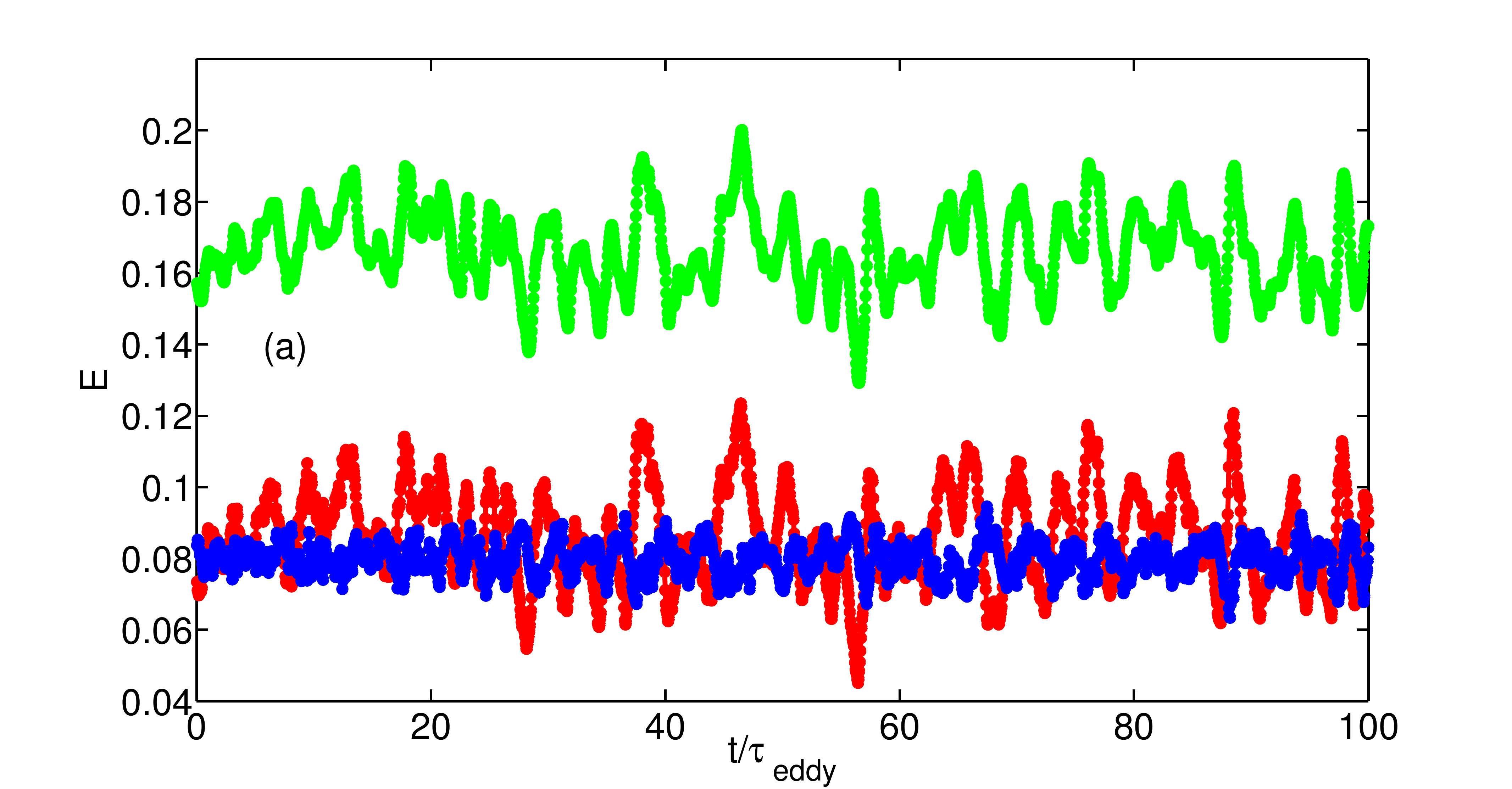}
\includegraphics[width=0.95\columnwidth]{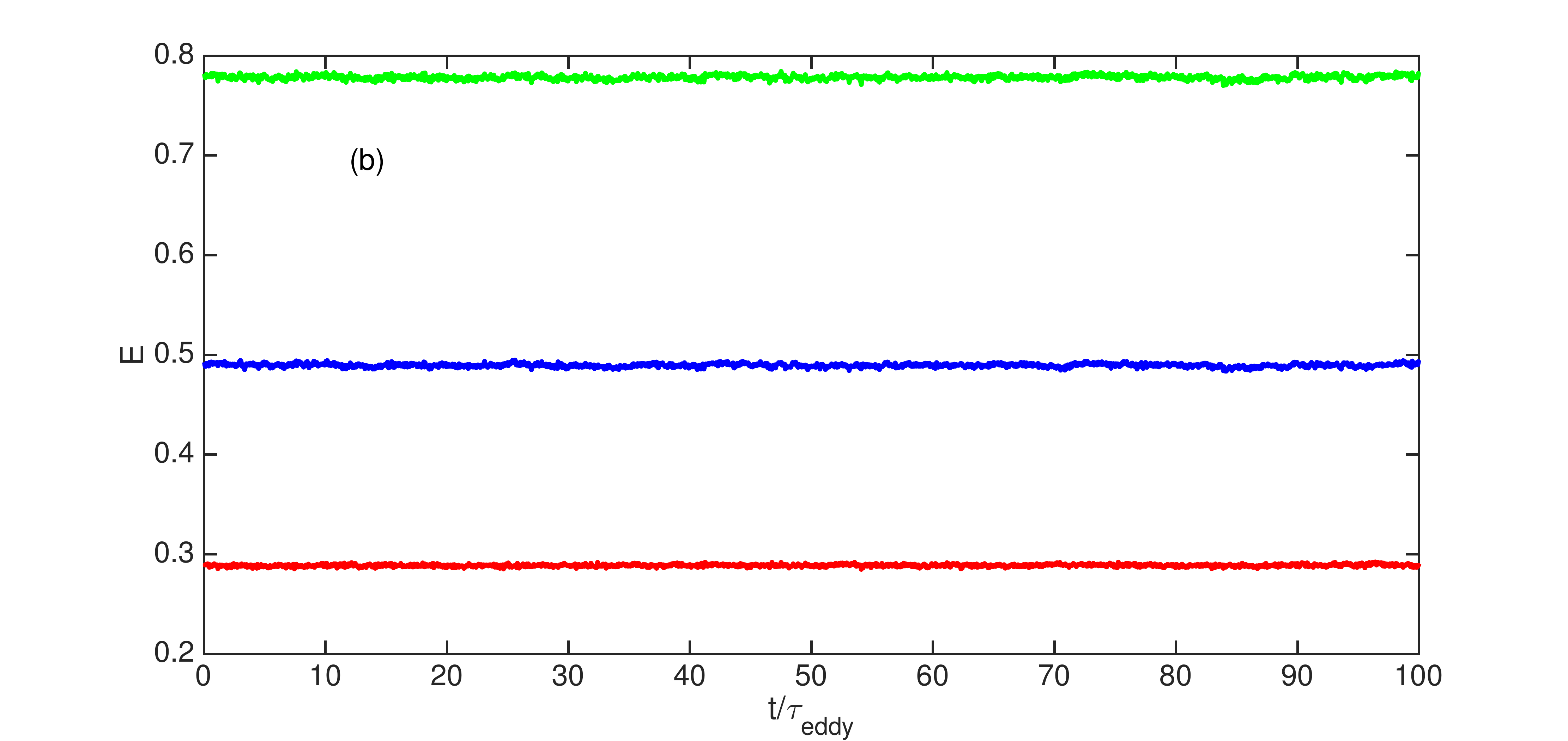}
\caption{Plots versus time of the kinetic (red curves), magnetic (blue curves),
and total (green curves) energies for runs (a) R1 and (b) R2. The origin on 
the horizontal axis is chosen at a time at which a statistically steady state 
has been obtained; data for averages are collected thereafter for $\simeq
100 \tau_{eddy}$. }
\label{fig:energy_time}
\end{figure}

In Figs.~\ref{fig:energy_time} (a) and (b) we show the time evolution of the
kinetic (red curve), magnetic (blue curve), and total (green curve) energies,
for runs (a) R1 and (b) R2, after the turbulent, nonequilibrium, statistically
steady states have been established by the forcing and dissipation terms in the
2D MHD equations. By comparing Figs.~\ref{fig:energy_time} (a) and (b), we see
that the properties of these statistically steady states are markedly different
for runs R1 (energy injection at a large length scale) and R2  (energy
injection at a small length scale): In the former case the kinetic energy and
magnetic energy are of the same magnitude $E^u \simeq E^b$; we refer to this
phenomenon as \emph{equipartition}; in the latter case, however, there is a clear
gap between the kinetic and magnetic energies and, at all values of $t/\tau_{\rm
eddy}$, we have $E^b > E^u$. The 2D MHD equations have a wide range of cascading
invariants.  Recent studies~\cite{Seshasayanan2014,Seshasayanan2016} have examined
the transition from hydrodynamic to MHD regimes in 2D MHD turbulence; this
transition takes place because of competing and counter-cascading
\emph{invariants} (of the ideal, 2D MHD equations). Our system, which is in the
MHD regime, has a forward-cascading energy and an inverse-cascading $|\psi|^2$.
In this regime, we propose the following dominant-balance argument to explain the 
lack of energy equipartition in our run R2: Energy is injected at the wavevector
$k_{\rm inj} = 250$, which corresponds to very small length scales.  The
forward-cascading energy is almost dissipated, locally, by the combined action of
the (scale-independent) friction and the hyperviscosity and
magnetic hyperdiffusivity (both dominant at very small
length scales). However, the inverse-cascading $|\psi|^2$ can go to larger
length scales, where the effect of hyperviscous dissipation is almost absent,
and the only dissipation mechanism is friction. Therefore, at large length
scales, we have kinetic energy mostly from the magnetic energy, which has
cascaded there, and not from the forcing in the equation of motion for the
velocity, whence we conclude that $E^u$ has to be less than $E^b$ in run R2.

\begin{figure}[htbp!]
\includegraphics[width=0.95\columnwidth]{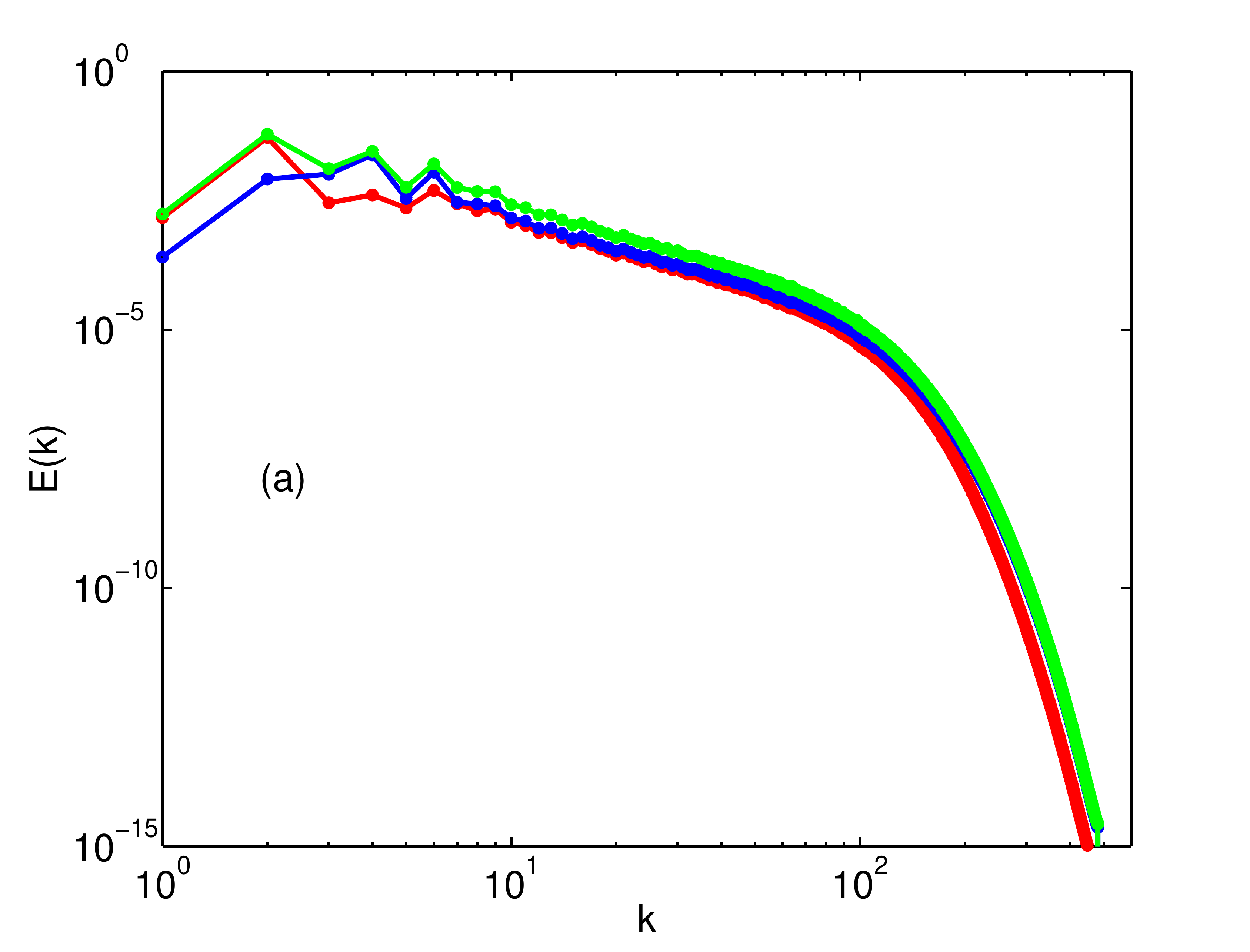}
\includegraphics[width=0.95\columnwidth]{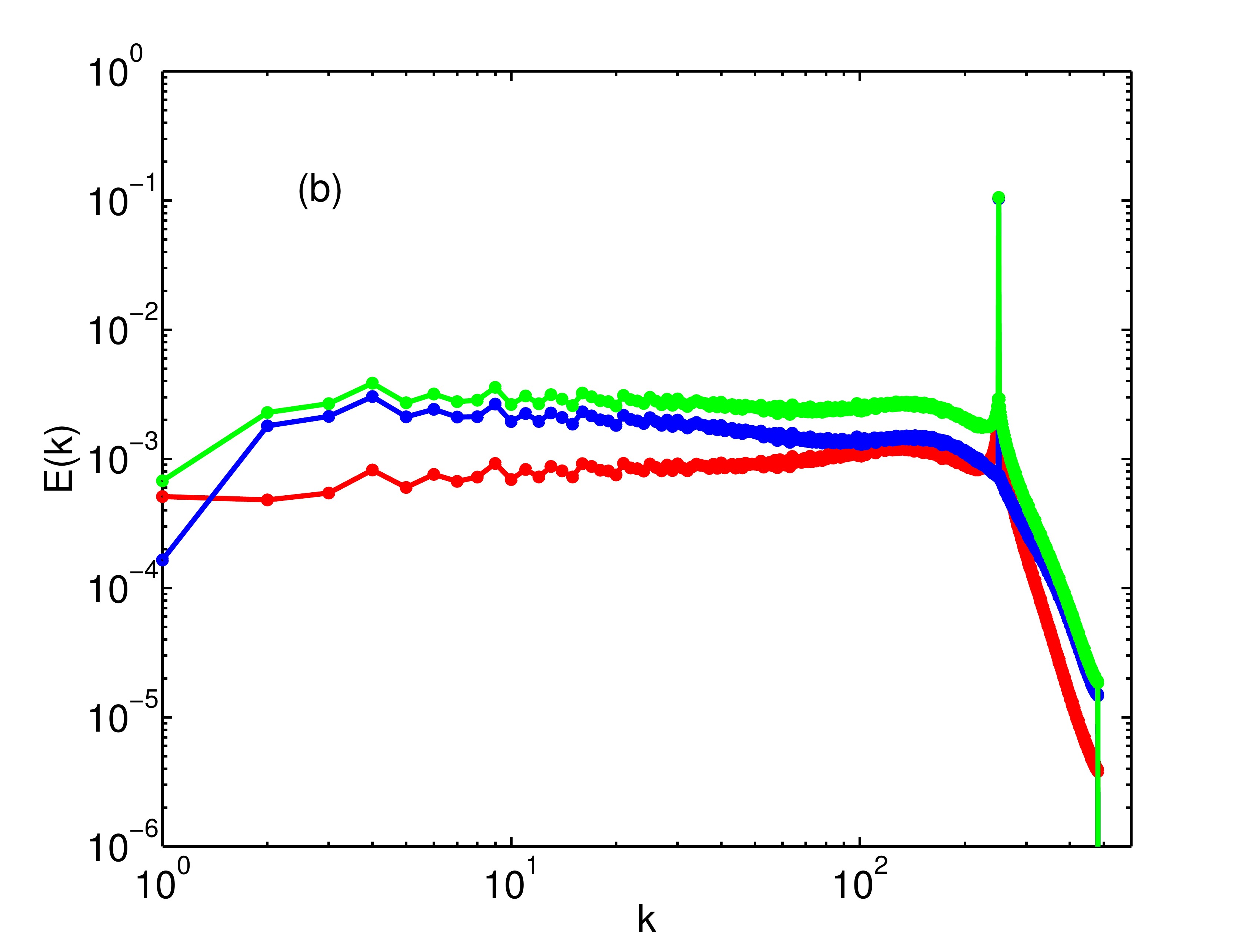}
\caption{Log-log plots versus the wave number $k$ of kinetic (red curves), 
magnetic (blue curves), and total (green curves) energy spectra, for the 
statistically steady states in runs (a) R1 and (b) R2.}
\label{fig:energy_spectra}
\end{figure}

In Figs.~\ref{fig:energy_spectra} (a) and (b) we give log-log plots, versus the
wave number $k$, of kinetic (red curves), magnetic (blue curves), and total
(green curves) energy spectra, for the statistically steady states in runs (a)
R1 and (b) R2. The former shows a substantial forward-cascade inertial range
with power-law scaling that is consistent with $E^u(k) \sim E^b(k) \sim
k^{-2}$; the latter shows clear, inverse-cascade scaling ranges with $E^u(k)
\sim k^{0.1}$ and $E^b(k) \sim k^{-0.2}$. The scaling exponents that we
have obtained from our DNSs are different from the dimensional predictions that we
have outlined in the Introduction. These differences in the
exponents arise principally because of the friction terms, which affect the 
velocity and magnetic fields at all length scales. Such friction-induced
modifications of energy-spectral exponents
have been reported previously in hydrodynamic 
turbulence (e.g., Refs.~\cite{perlekarnjp,pramanareview09} and references therein).
Note also that in both Figs.~\ref{fig:energy_spectra} (a) and (b) the energy spectra
fall at small $k$ (near the $k=1$ mode) because of the friction terms,
which generate a small-$k$ cutoff in the energy spectra~\cite{perlekarnjp,pramanareview09}.

%
%

\begin{figure}[htbp!]
\includegraphics[width=0.95\columnwidth]{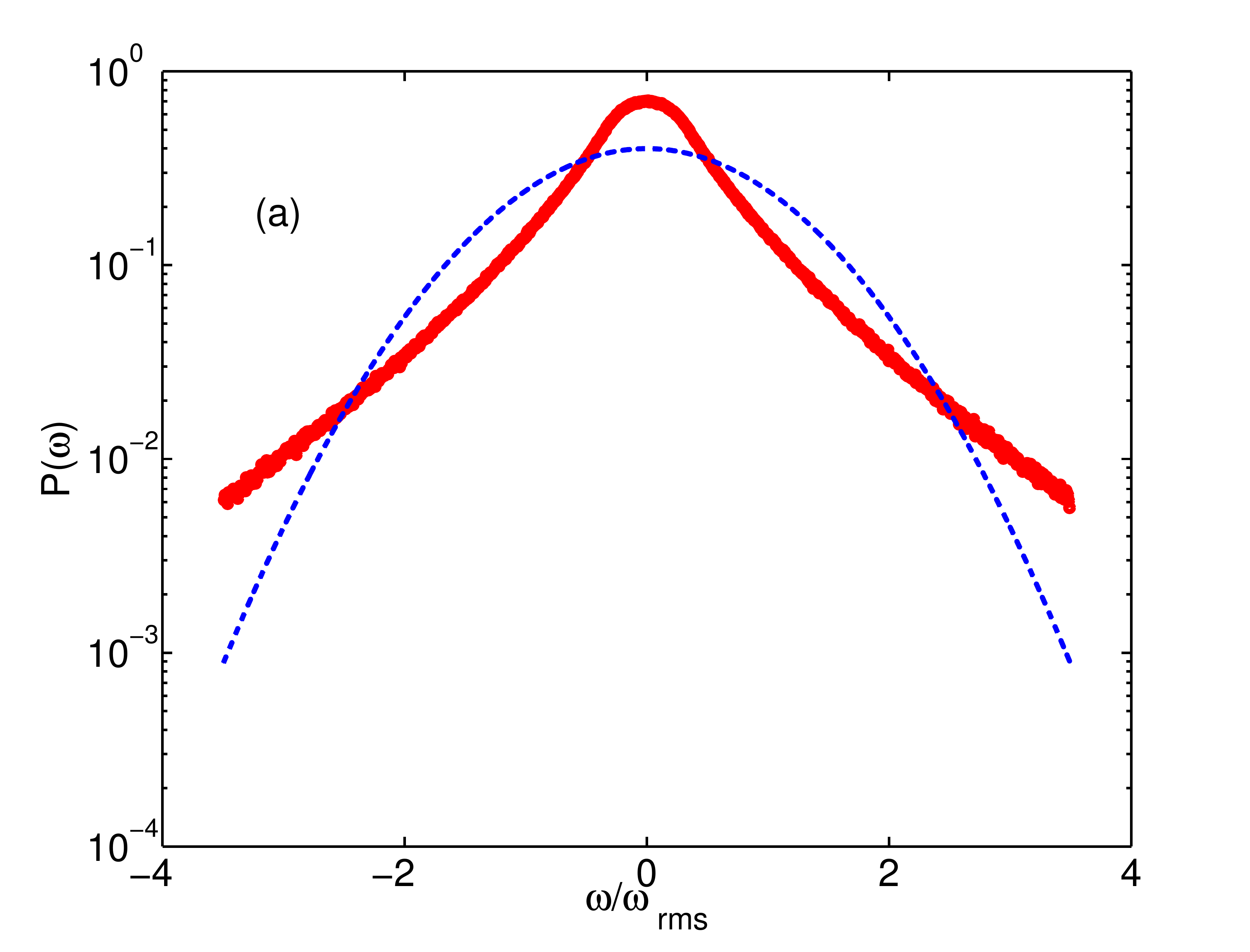}
\includegraphics[width=0.95\columnwidth]{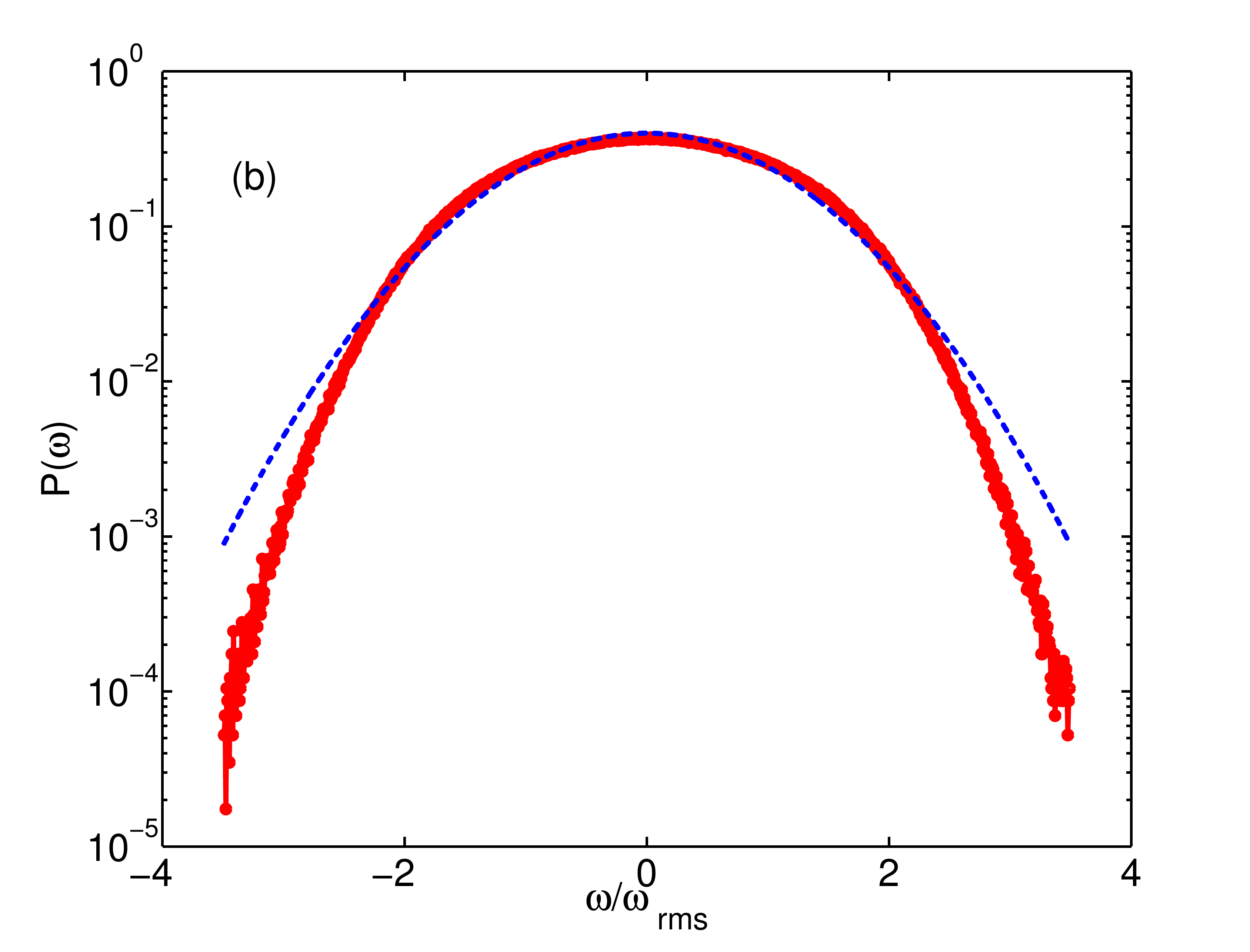}
\caption{Semilogarithmic plots the PDF $P(\omega)$ of the vorticity $\omega$
for the runs (a) R1 and (b) R2. The blue, dashed curves indicate Gaussian
distributions for comparison.}
\label{fig:vorticity_pdf}
\end{figure}

\begin{figure}[htbp!]
\includegraphics[width=0.95\columnwidth]{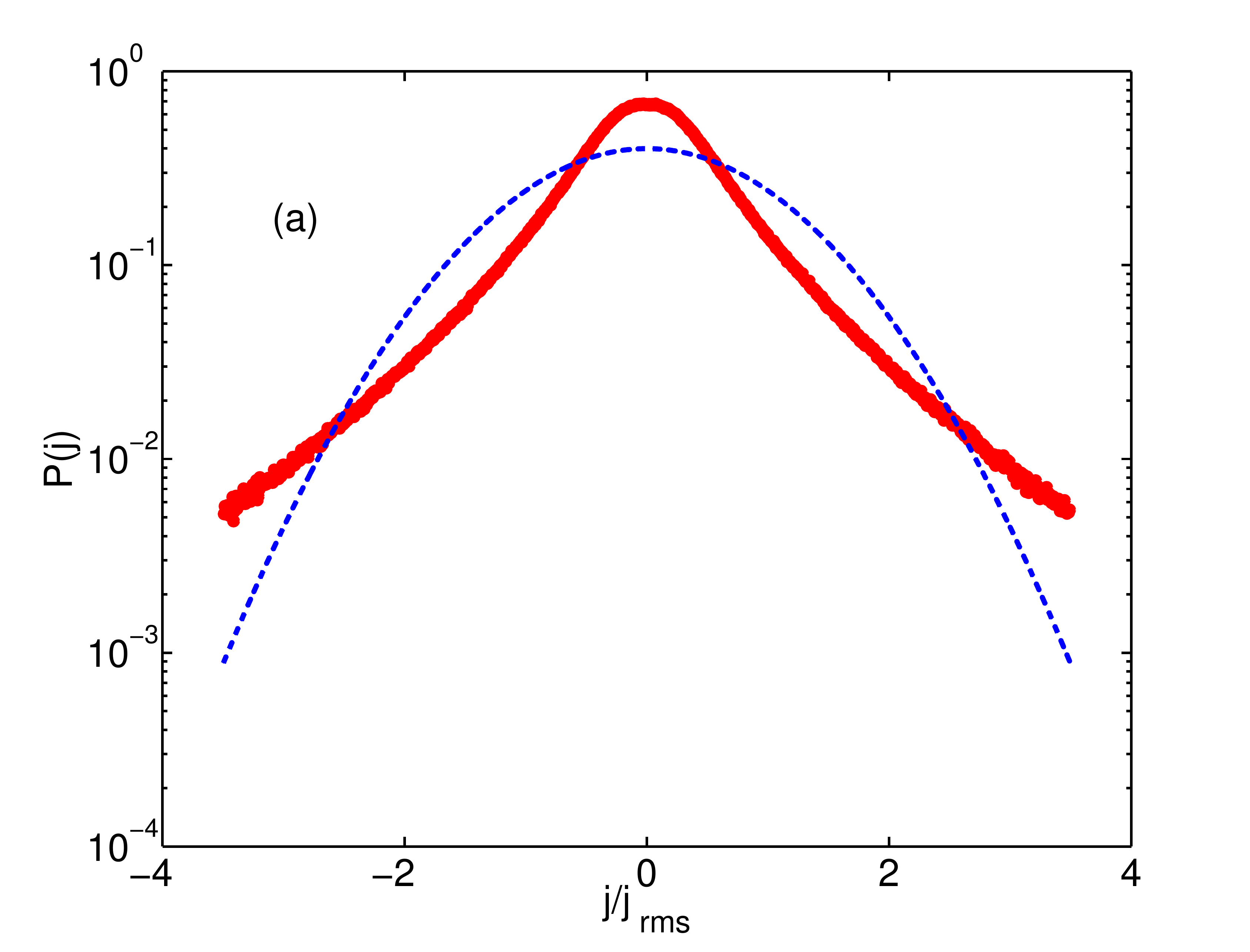}
\includegraphics[width=0.95\columnwidth]{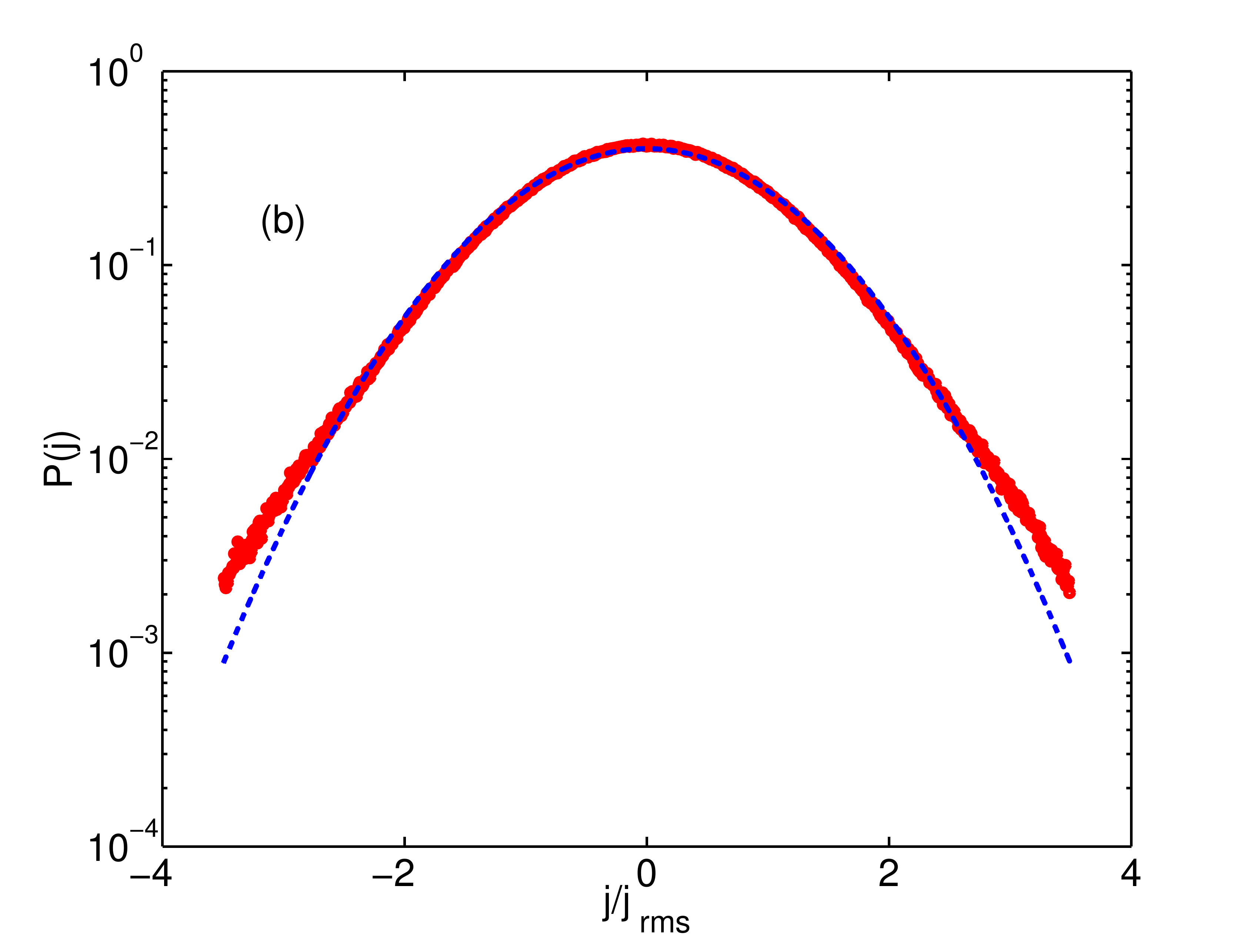}
\caption{Semilogarithmic plots the PDF $P(j)$ of the current density $j$
for the runs (a) R1 and (b) R2. The blue, dashed curves indicate Gaussian
distributions for comparison.}
\label{fig:current_pdf}
\end{figure}

In Figs.~\ref{fig:vorticity_pdf} (a) and (b) we plot the PDFs $P(\omega)$ of
$\omega$ for runs R1 and R2, respectively. The PDF of $\omega$ (run R1)
deviates significantly from the Gaussian distribution, denoted by the blue
dashed curve.  The probability of large vales of $\omega$ is high compared to
what we expect from a Gaussian PDF. By contrast, in run R2, this PDF is much
closer to a Gaussian, and the tail of the PDF is sub-Gaussian. This
is consistent with the earlier observation (e.g.,
Ref.~\cite{Boffetta2000,Bernard2007}) that the inverse-cascade regime is scale
invariant, whereas the forward-cascade regime is associated with
\emph{intermittency}.  In Figs.~\ref{fig:current_pdf} (a) and (b) we plot the
PDFs of $j$ for runs R1 and R2, respectively. The former is distinctly
non-Gaussian, but the latter is close to a Gaussian PDF. However, there is a
weak super-Gaussian tail in the PDF of $j$ (for run R2), which is unlike its
vorticity counterpart.
 
\begin{figure}[htbp!]
\includegraphics[width=0.95\columnwidth]{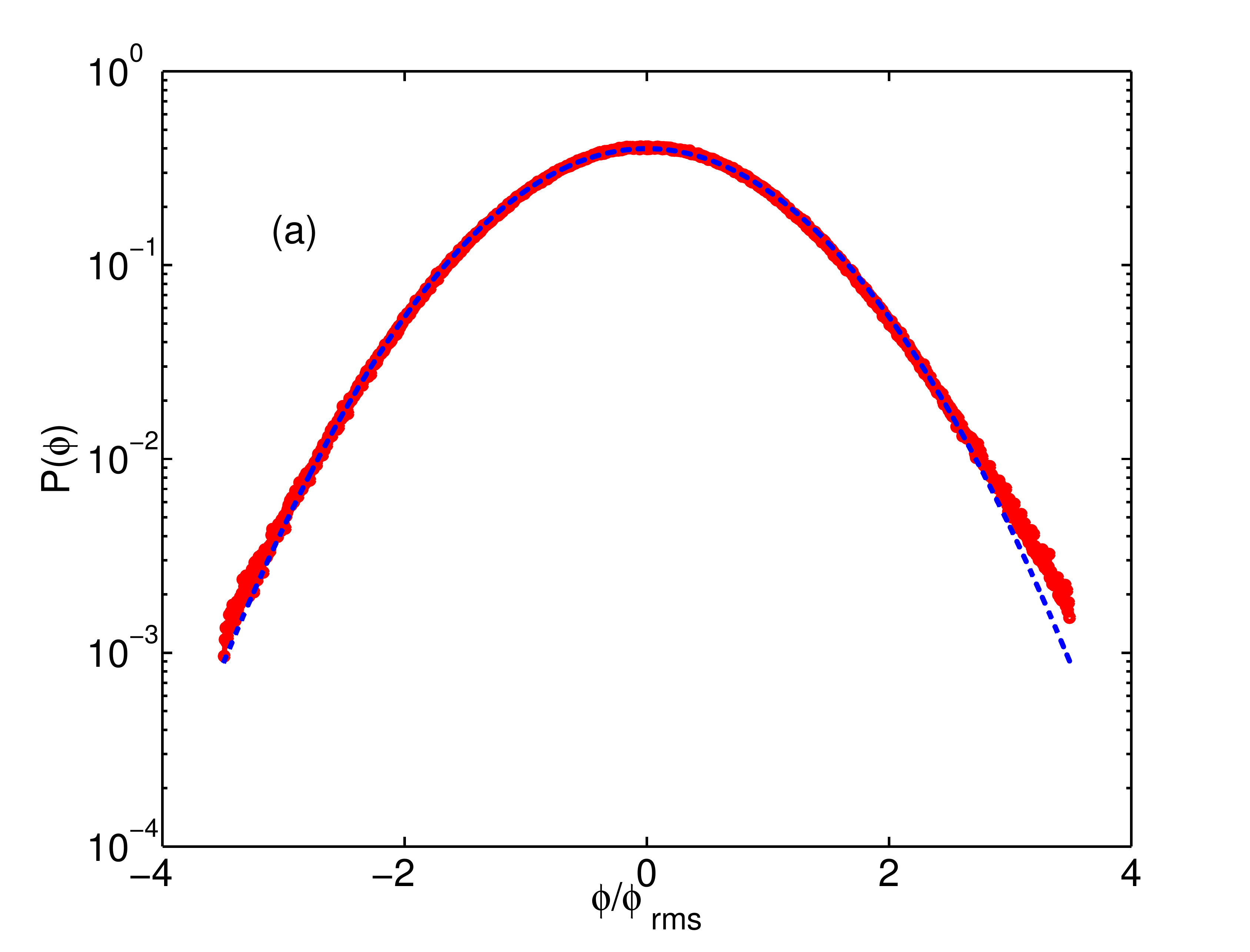}
\includegraphics[width=0.95\columnwidth]{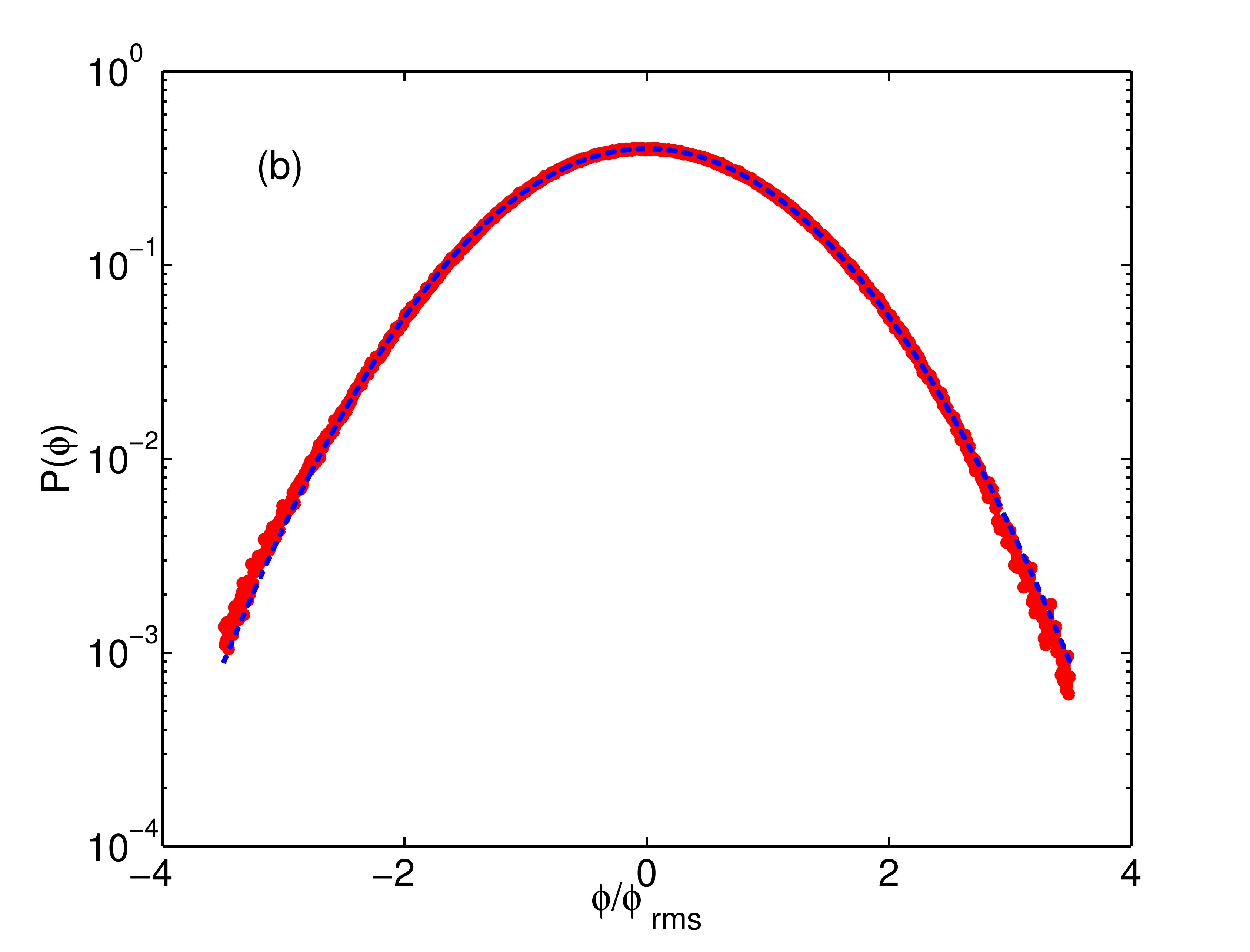}
\caption{Semilogarithmic plots the PDF $P(\phi)$ of the fluid stream
function $\phi$ for the runs (a) R1 and (b) R2. The blue, dashed curves 
indicate Gaussian distributions for comparison.}
\label{fig:phi_pdf}
\end{figure}

\begin{figure}[htbp!]
\includegraphics[width=0.95\columnwidth]{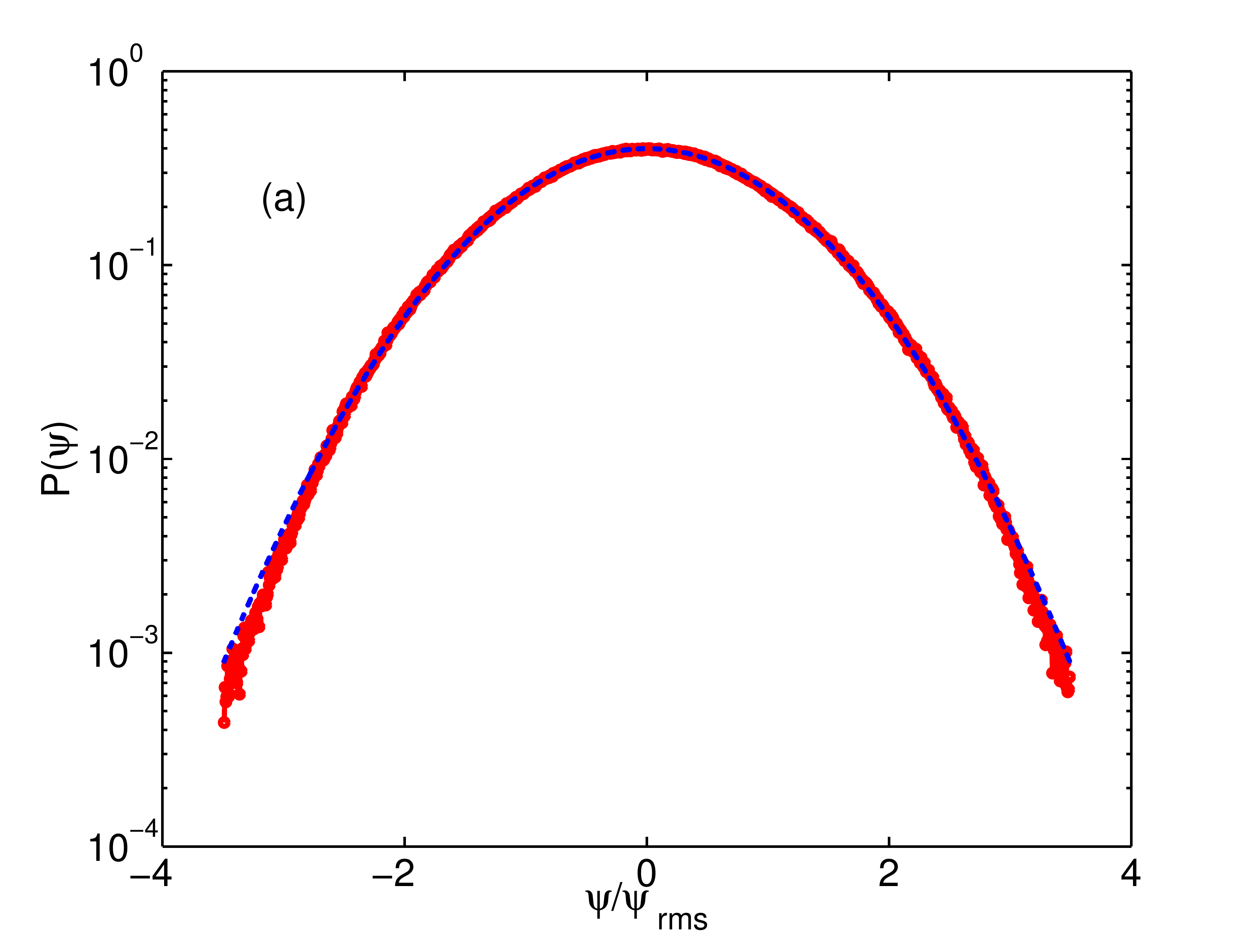}
\includegraphics[width=0.95\columnwidth]{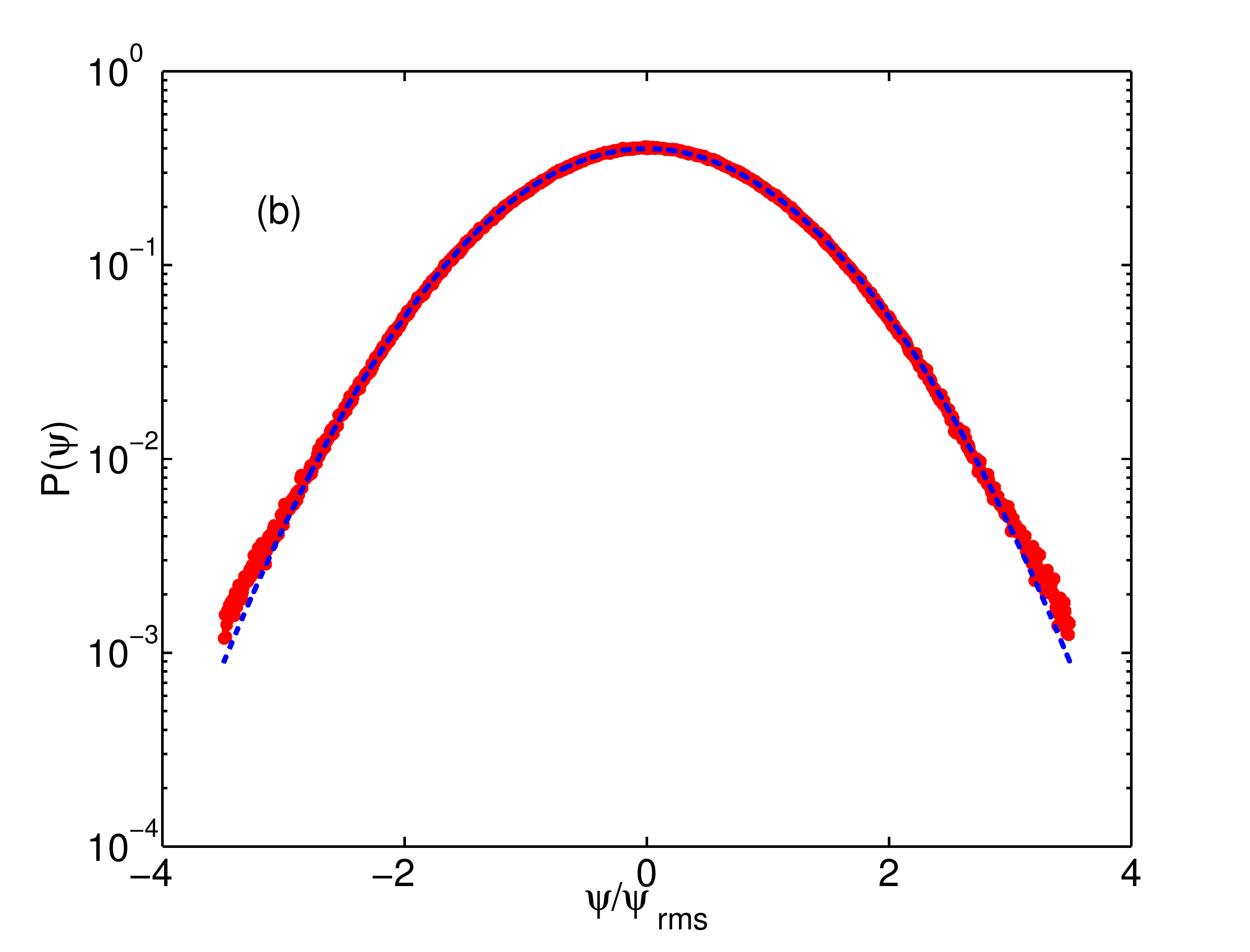}
\caption{Semilogarithmic plots the PDF $P(\psi)$ of the magnetic
potential $\psi$ for the runs (a) R1 and (b) R2. The blue, dashed curves 
indicate Gaussian distributions for comparison.}
\label{fig:psi_pdf}
\end{figure}

In Figs.~\ref{fig:phi_pdf} (a) and (b) we plot the PDFs $P(\phi)$ of the fluid
stream function $\phi$, for runs R1 and R2, respectively. Their counterparts
$P(\psi)$, for the magnetic potential $\psi$, are shown in
Figs.~\ref{fig:psi_pdf} (a) and (b). All these PDFs are almost Gaussian, with
very-small deviations in their tails. This observation has implications for
intermittency in the forward-cascade regime. The vorticity and the current
density, which show strong non-Gaussian PDFs in run R1, are second spatial
derivatives of the stream function and the magnetic potential, respectively.
The higher the order of the spatial derivatives the smaller the length scales 
at which these derivatives contribute significantly: these are the small length scales at
which we obtain intermittency in 2D MHD turbulence.

\begin{figure}[htbp!]
\includegraphics[width=0.95\columnwidth]{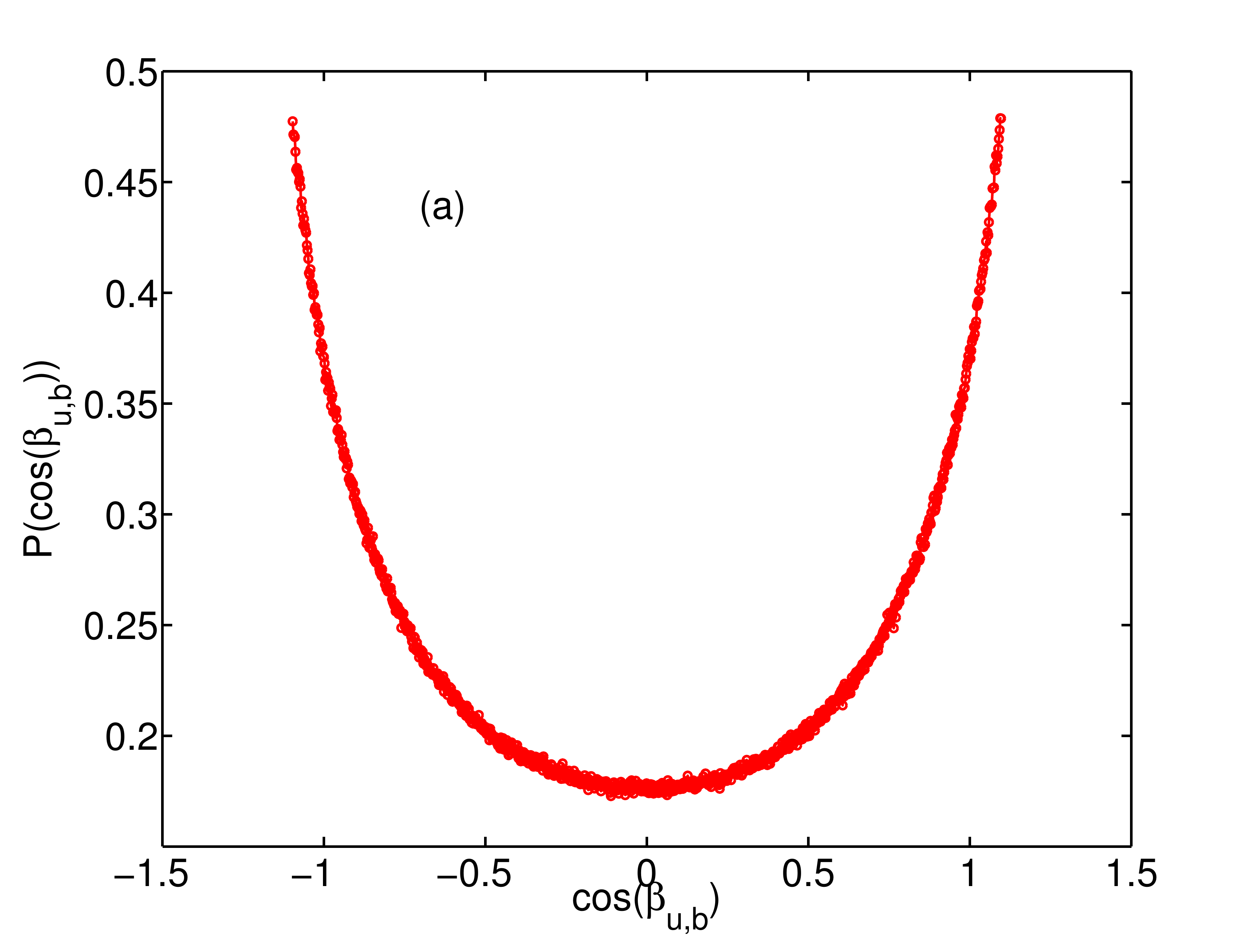}
\includegraphics[width=0.95\columnwidth]{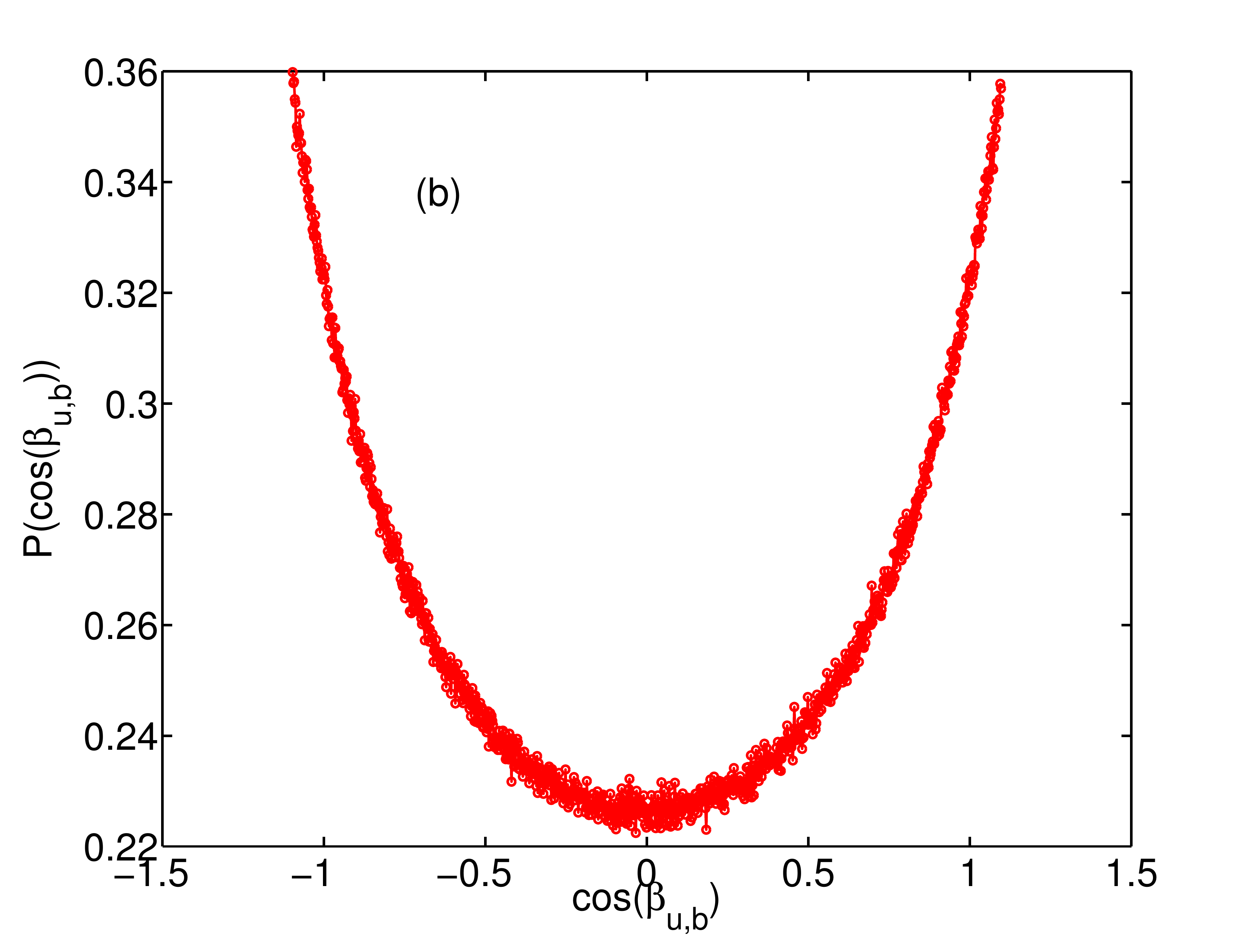}
\caption{Plots of the PDFs of the cosine of the angle $\beta_{u,b}$
between the velocity field $\bf u$ and the magnetic field
$\bf b$ for runs (a) R1 and (b) R2.}
\label{fig:align}
\end{figure}

We now explore the alignment of ${\bf u}$ and ${\bf b}$ by plotting, in
Figs.~\ref{fig:align} (a) and (b), the PDFs of the cosine of the angle
$\beta_{u,b}$ between the velocity and magnetic fields for runs (a) R1 and (b)
R2, respectively. In both these cases, these PDFs show that there is a
significant tendency for  ${\bf u}$ and ${\bf b}$ to be aligned or
anti-aligned. In run R1 (forward-cascade domination) this tendency is
greater than in run R2 (inverse-cascade domination). The typical
probability for the vectors ${\bf u}$ and ${\bf b}$ to be orthogonal to
each other is $\simeq 20\%$; although this is smaller
than the probability of having aligned and anti-aligned states, non-aligned
states play a very important role in 2D MHD. To understand
this, consider the induction equation:
\begin{equation}
\frac{\partial {\bf b}}{\partial t} = \nabla \times \left( {\bf u} \times {\bf b} \right) + \eta \nabla^2 {\bf b};
\end{equation}
the nonlinear term on the right-hand side is identically zero for perfectly
aligned or anti-aligned ${\bf u}$ and ${\bf b}$, in which case this induction
equation reduces to the linear diffusion equation for the magnetic field
only~\cite{Banerjee2014,marino}. To the extent that 2D MHD turbulence does
not display purely diffusive behavior, the alignment-induced depletion of
nonlinearity is not complete.

\begin{figure}[htbp!]
\includegraphics[width=0.95\columnwidth]{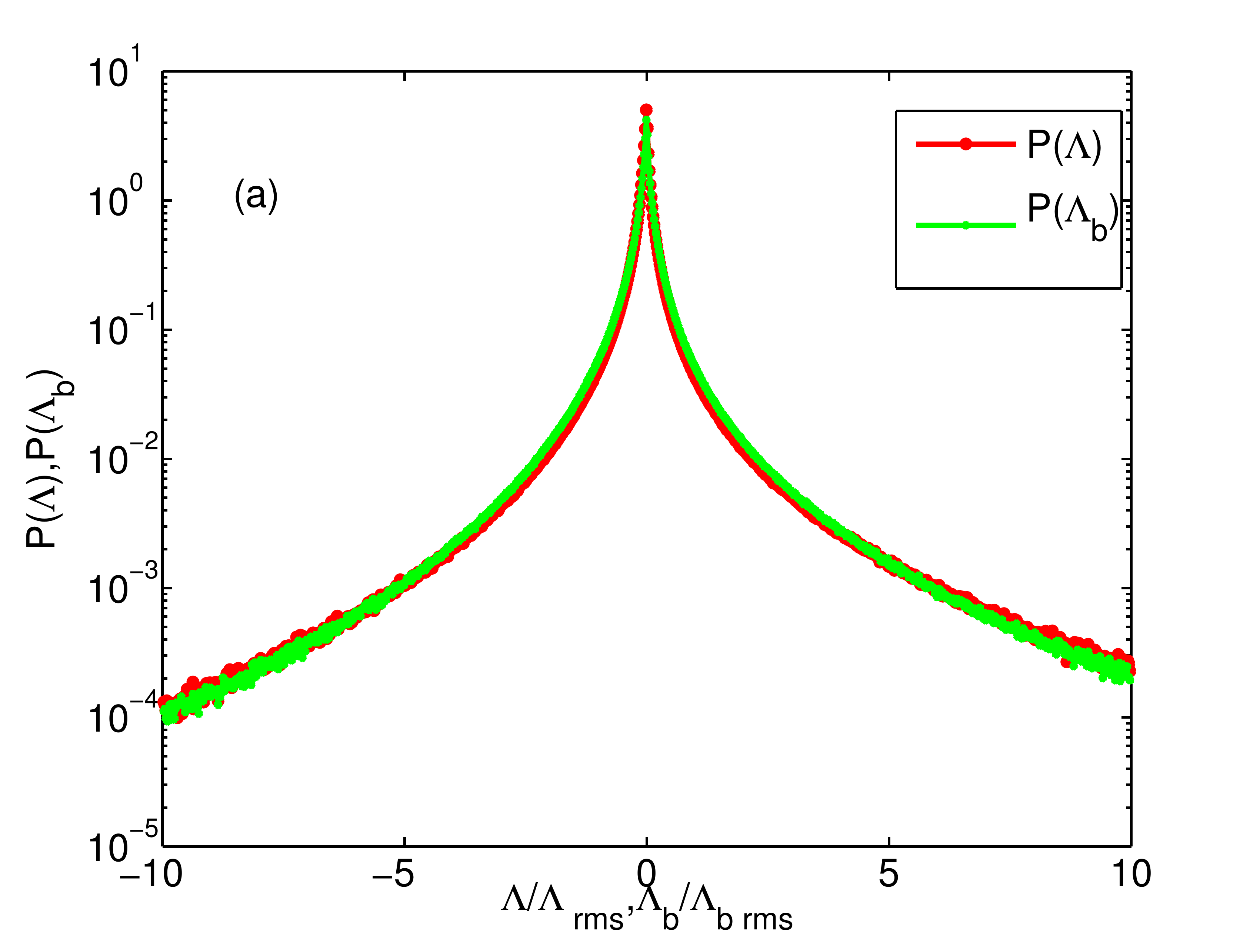}
\includegraphics[width=0.95\columnwidth]{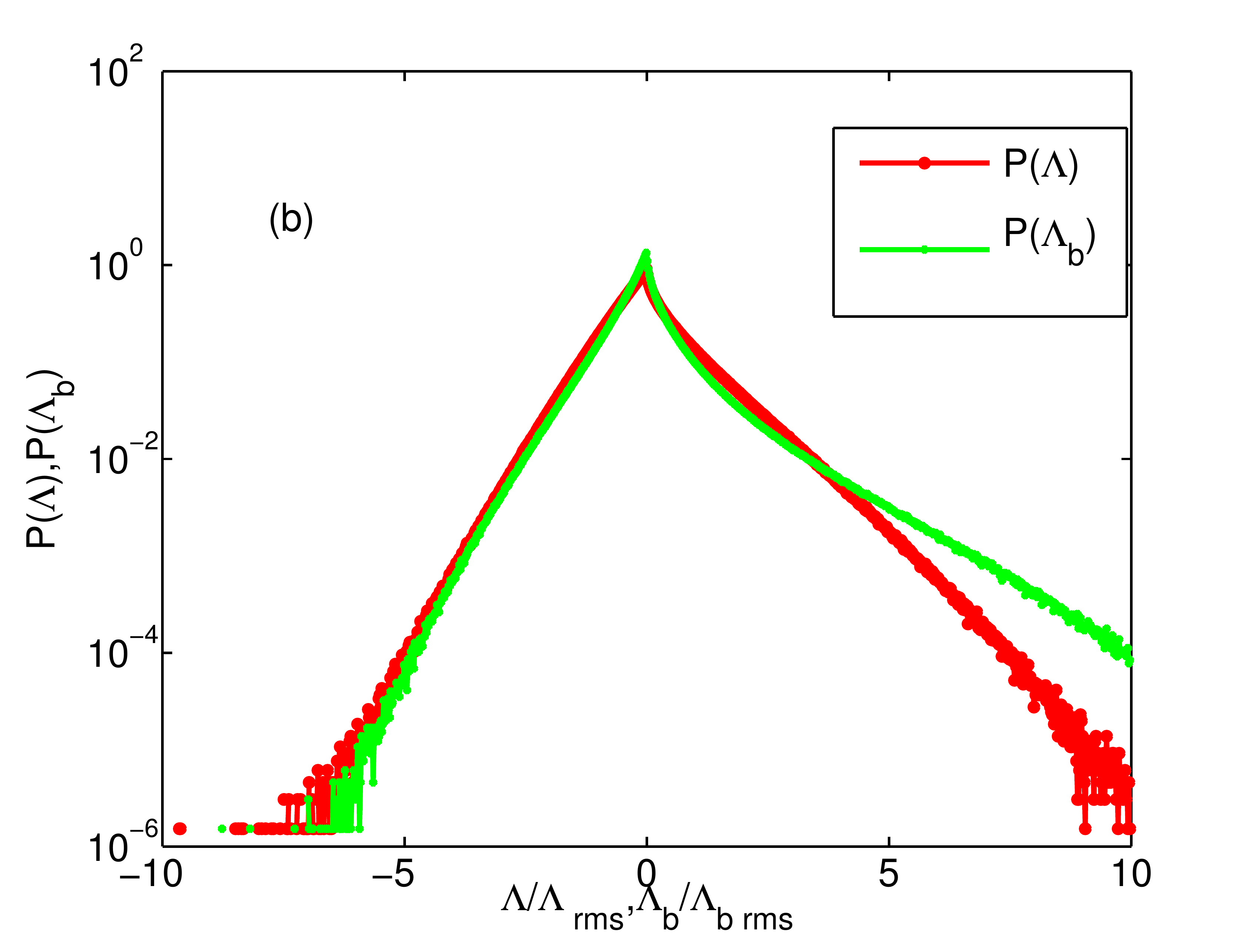}
\caption{Semilogarithmic plots of the PDFs of the Okubo-Weiss parameter 
$\Lambda$ and its magnetic counterpart $\Lambda_b$ for the runs (a) R1 
and (b) R2; the subscript $rms$ stands for root-mean square.}
\label{fig:pdf_okw}
\end{figure}

In Figs.~\ref{fig:pdf_okw} (a) and (b) we plot, for runs R1 and R2, 
respectively, the PDFs of the Okubo-Weiss parameter $\Lambda$ and its 
magnetic analogue $\Lambda_b$, which are~\cite{Banerjee2014,shivamoggi}
\begin{eqnarray}
\Lambda  &=& -(\partial u_x/\partial x)^2  - (\partial u_y/\partial x) 
(\partial u_x/\partial y) ; \nonumber \\ 
\Lambda_b  &=& -(\partial b_x/\partial x)^2  - (\partial b_y/\partial x) 
(\partial b_x/\partial y) .
\label{eq:Lambda}
\end{eqnarray}
For a fluid, in the inviscid, unforced case without friction, the sign of
$\Lambda$ can be used to distinguish between vortical ($\Lambda > 0$) and
extensional regions ($\Lambda < 0$) regions of the
flow~\cite{okubo,weiss,Banerjee2014,perlekarnjp}.  This criterion works well
even in the presence of viscosity, friction, and forcing~\cite{perlekarnjp}.
The magnetic analog $\Lambda_b$ of the fluid Okubo-Weiss parameter is positive
in current-dominated regions and negative in regions that are dominated by the
magnetic strain rate~\cite{Banerjee2014}. Figures~\ref{fig:pdf_okw} (a) and (b)
show that the PDFs $P(\Lambda)$ and $P(\Lambda_b)$ show cusps at $\Lambda = 0$
and $\Lambda_b = 0$, respectively, and have distinctly non-Gaussian tails;
these tails are broader for run R1 than for run R2, which signifies again that
extreme events and intermittency are more probable in the forward-cascade case
(run R1) than in the inverse-cascade one (run R2). 

\section{Conclusions}
\label{sec:concl}

We have presented a DNS study of 2D, homogeneous, isotropic MHD turbulence.  In
particular, we have compared the statistical properties of such turbulence for
our DNS runs R1 and R2, in which we obtain statistically steady states with forcing such
that $k_{\rm inj}=2$ and $k_{\rm inj}=250$. We have shown that the statistical
properties of the turbulent states, in runs R1 and R2, are strikingly
different. We have demonstrated this by calculating  and comparing, for these two
runs, (a) the time evolution of the kinetic, magnetic, and total energies, (b)
energy spectra and fluxes, and (c) PDFs of the
vorticity, current density, fluid stream function, magnetic potential, of the
cosine of the angle between the velocity and magnetic fields, and of the
Okubo-Weiss parameter~\cite{okubo,weiss} and its magnetic
analog~\cite{shivamoggi,Banerjee2014}, which help us to characterise the
topology of the flow. We have demonstrated, \textit{inter alia}, that the
probability of extreme events, characterised, say, by large values of $\omega,
\, j, \Lambda,$ and $\Lambda_b$, is higher in run R1 than in run R2. 
We hope our study will lead to similar, systematic comparisons of the statistical
properties of turbulence in systems that exhibit both forward and inverse cascades.

%
%
%
%

\begin{acknowledgements}

DB thanks the Cost Action MP 1305 for support. RP thanks DST, CSIR,
and UGC India for support and SERC (IISc) for computational resources.

\end{acknowledgements}

\end{document}